# Status of the HENSA collaboration at the Canfranc Underground Laboratory: results from two years of measurement of the neutron flux in hall B

**N. Mont-Geli,**[a,*] **A. Tarifeño-Saldivia,**[b] **G. Cortés,**[a] **J. L. Tain,**[b] **M. Grieger,**[c] **A. Quero-Ballesteros,**[d] **M. Pallàs,**[a] **J. Agramunt,**[b] **A. Algora,**[b] **J. Amaré,**[g] **D. Bemmerer,**[c] **F. Calviño,**[a] **D. Cano-Ott,**[i] **S. Cebrián,**[g] **D. Cintas,**[g] **I. Coarasa,**[g] **A. De Blas,**[a] **I. Dillman,**[f] **L. M. Fraile,**[h] **E. García,**[g] **R. Garcia,**[a] **A. M. Lallena,**[d] **M. Martínez,**[g] **T. Martínez,**[i] **E. Nacher,**[b] **S. E. A. Orrigo,**[b] **Y. Ortigoza,**[g] **A. Ortiz de Solórzano,**[g] **T. Pardo,**[g] **J. Plaza,**[i] **J. Puimedón**[g] **and M. L. Sarsa**[g]

[a]*Institut de Tècniques Energètiques (INTE), Universitat Politècnica de Catalunya (UPC),
Av. Diagonal 647, Barcelona, Spain*

[b]*Instituto de Física Corpuscular (IFIC), CSIC - Univ. Valencia,
C. Catedrático José Beltrán 2, Paterna, Spain*

[c]*Helmholtz-Zentrum Dresden-Rossendorf (HZDR),
Bautzner Landstraße 400, Dresden, Germany*

[d]*Dpto. Física Atómica, Molecular y Nuclear, Facultad de Ciencias, Universidad de Granada,
Campus Fuente Nueva s/n, Granada, Spain*

[f]*TRIUMF,
4004 Westbrook Mall, Vancouver, Canada*

[g]*Centro de Astropartículas y Física de Altas Energías (CAPA), Universidad de Zaragoza,
C. Pedro Cerbuna 12, Zaragoza, Spain*

[h]*Grupo de Física Nuclear & IPARCOS, Universidad Complutense de Madrid (UCM),
Pl. de las Ciencias 1, Madrid, Spain*

[i]*Centro de Investigaciones Energéticas, Medioambientales y Tecnológicas (CIEMAT),
Av. Complutense 40, Madrid, Spain*

*E-mail:* nil.mont@upc.edu, atarisal@ific.uv.es

This work deals with the characterization of the neutron flux in hall B of the Canfranc Underground Laboratory (LSC) employing the High Efficiency Neutron Spectrometry Array (HENSA). The ultimate goal of this measurement is to set a limit on the corresponding effects of the neutron flux in the background of the ANAIS-112 experiment. The preliminary neutron counting rates of two years of measurement are reported. Various data analysis techniques, including pulse shape discrimination, are discussed. The first results on the spectral reconstruction of the neutron flux are also presented.

[*]Speaker

**1. Introduction**

The Canfranc Underground Laboratory (LSC) is an underground facility located 850 m under Mount Tobazo in the Aragonese Pyrenees [1], thus providing a very low-background environment suitable for rare-event measurements in multiple fields [2]. However, even underground, the neutron flux could be large enough to be one of the main limiting factors in such type of experiments. A detailed characterization of the neutron flux is, therefore, required.

The High Efficiency Neutron Spectrometry Array (HENSA) [3] is a detection system based on the same principles as Bonner Spheres (BS) spectrometers [4]. It consists of ten long $^3$He-filled neutron proportional counters which are embedded in blocks of High Density PolyEthylene (HDPE) with different sizes in order to provide sensitivity in a wide range of neutron energies. The use of long tubes (60 cm length) provides HENSA with a neutron detection efficiency around one order of magnitude larger than standard BS spectrometers. Early versions of HENSA have already been used for the assessment of the neutron flux in hall A of the LSC [5–7] and in the shallow underground facility Felsenkeller in Dresden (Germany) [8].

This work deals with the characterization of the neutron flux in hall B of the LSC. The main motivation is to provide information about the neutron flux that could affect the background of ANAIS-112, an experiment looking for the expected annual modulation of the Dark-Matter signal in NaI scintillators [9].

**2. Simulations of the neutron flux**

Monte Carlo calculations of the neutron flux has been carried out in order to understand the neutron sources and to provide prior information for the unfolding algorithms. These calculations are based on the previous work by M. Grieger *et al.* [8]. Three different types of neutron sources has been evaluated: ($\alpha$,n) reactions, spontaneous fission and ($\mu$,n) reactions [10]. For the radiogenic sources only the natural decay chains from the $^{232}$Th and $^{238}$U series has been considered.

Key inputs for the simulation are the chemical compositions and the intrinsic activities of the Canfranc rock and concrete walls. Although a detailed characterization of the rock exists [11], there is no any information about the composition of the concrete walls. Consequently, different types of standard concretes [12] have been used. Information about the concrete radioactivity has been provided by the LSC staff.

Neutron transport calculations are carried out using FLUKA [13–15]. The Geant4 tool SaG4N [16, 17] is employed to determine the ($\alpha$,n) production yields and spectra. The spontaneous fission yields are derived from the measured activities and the energy spectrum is assumed to be described by a Watt spectrum [18]. The muon-induced neutron production is directly computed using the FLUKA code. The muon flux has been recently measured [19] and a depth-dependent parameterization [20] is used for the energy distribution.

Results are shown in figure 1. From thermal up to fast energies (below 20 MeV) the main contribution are neutrons produced within the concrete walls. In the high-energy region (above 20 MeV) muon-induced neutrons are the only contribution. It must be kept in mind that the calculations have been carried out assuming the geometry of hall A. Nevertheless, except for the magnitude of the integral flux, no relevant differences are expected in respect of hall B.

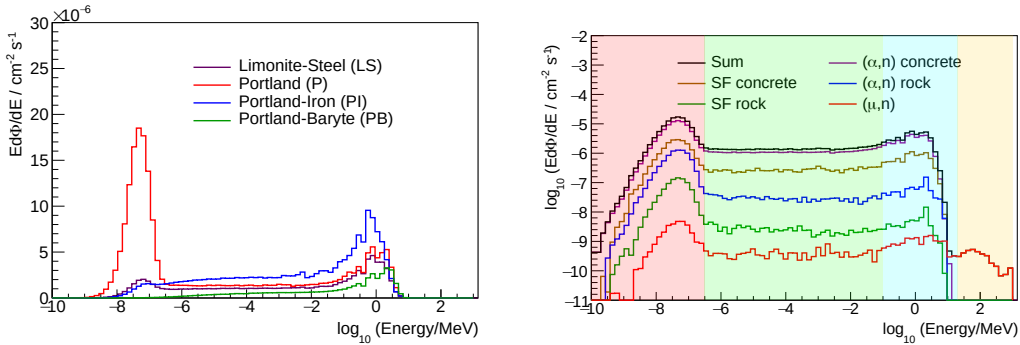

**Figure 1:** Monte Carlo calculations of the neutron flux in hall A of the LSC. Left: effect of using different types of concrete. Right: contribution of each neutron source compared to the total flux (assuming Portland concrete). Colors represent different region of the energy spectrum: thermal (red), intermediate (green), fast (blue) and high-energy (orange).

## 3. The HENSA setup in hall B of the LSC

The experimental campaign involved three different setups: a reduced version using only three detectors to monitor different regions of the spectrum [21], the complete setup in the version (v2019) previously used in hall A [6, 7] and a new version with improved spectral resolution up to 20 MeV (v2022). The main difference between both setups (see figure 2) is the removal of the high-energy detectors in v2019 and their replacement for detectors which are sensible to lower energy neutrons. Consequently, with the new setup we are not able to measure muon-induced neutrons. Anyway, due to a lack of efficiency, it was also not possible to measure such type of neutrons with the old setup.

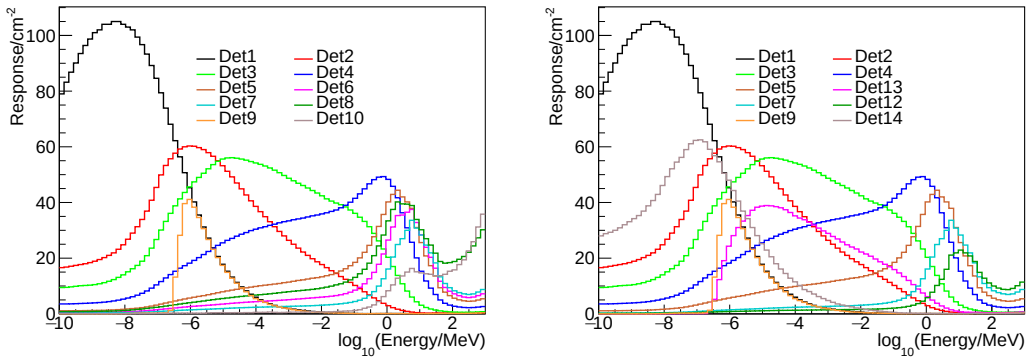

**Figure 2:** Response matrices of both versions of the HENSA setup in hall B: v2019 (left) and v2022 (right). Calculations have been carried out using the Geant4 tool Particle Counter [22].

## 4. Data analysis: neutron counting rates

Figure 3 shows the energy deposition spectra in the $^3$He counters of HENSA. The contribution of neutrons is overlapped with the contribution of other events such as electronic noise, micro-discharges, gamma-rays and alpha-particles from the intrinsic radioactivity of the counter walls. The latter is the sole responsible for the high energy component above 1000 keV.

The standard analysis methodology [5, 7, 21] is based on the use of a $^{252}$Cf source in order to obtain a "pure" neutron spectrum which summed to the alpha background is scaled to the "raw" data (red). The scale factor is computed assuming that there is no other relevant contribution in the region of the 765 keV peak. However, in some periods the contribution of noise events was so high (right) that a beforehand filtering of such events was required (see next paragraph). The area under the scaled spectrum (green line) after the subtraction of the alpha contribution is considered equivalent to the number of neutron events. The estimation of the alpha background is based on the assumption that the linear shape of the high energy region can be extrapolated to lower energies.

An alternative method is the use of Pulse Shape Discrimination (PSD) techniques or filters to remove non-neutron events from the energy deposition spectrum. The effect of using different PSD filters is shown in different colors in figure 3. Noise and micro-discharge events are easily distinguished due to a clearly differentiate shape (figure 3, right). In the rise-time versus pulse-height space it is possible to separate gamma-rays from neutron and alpha events [23]. The alpha background is treated in the same way as in the standard method.

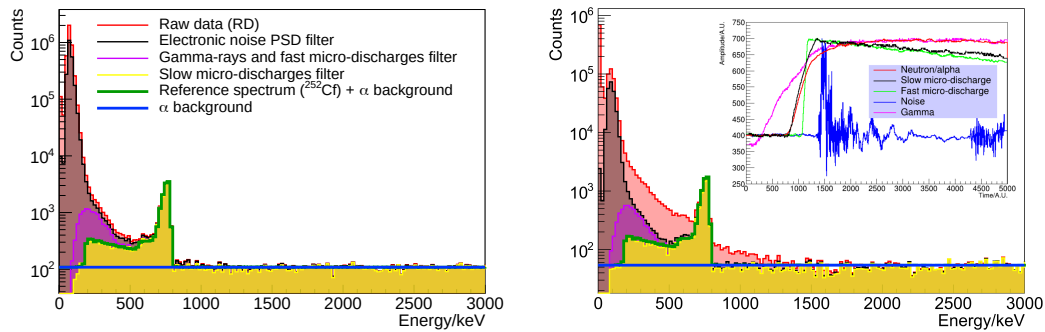

**Figure 3:** Typical energy deposition spectra in the $^3$He counters of HENSA. The effect of each PSD filter is shown in different colors. Left: low noise measurement. Right: high noise measurement.

The neutron rates (figure 4, preliminary results) are almost stable within the statistical uncertainties. Both methods provide consistent results.

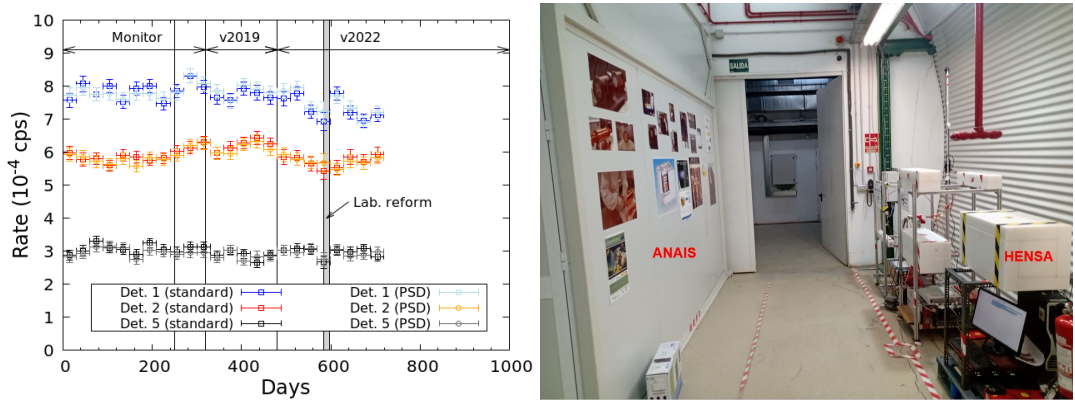

**Figure 4:** Left: preliminary neutron rates in the three detectors of the monitor setup from early April 2021 until late March 2023. Each phase is separated by black lines. Right: HENSA setup in hall B (v2022).

## 5. Data analysis: neutron energy distribution

The neutron flux has been reconstructed from the counting rates and the response matrix using three different unfolding algorithms: MAXED [24], GRAVEL [25] and a method based on the Bayes theorem [26]. Monte Carlo calculations from section 2, ignoring the muon contribution, served for defining the initial energy distribution feeding the unfolding algorithms (prior information).

Preliminary results using the average rates from phase 3 (HENSA-v2022, 168 life-days) are presented in figure 5. The red spectrum is the result of averaging a set of unfolded spectra obtained using the same prior (Portland concrete) and different methods (MAXED, GRAVEL and Bayes theorem). The blue spectrum is the result of averaging a set of unfolded spectra obtained using different priors and the Bayesian method. The most relevant differences are found in the region of thermal peak. The error-bars are the standard deviations.

The integral magnitude ($\phi_B = 20.4_1 \cdot 10^{-6}$ cm$^{-2}$ s$^{-1}$) is larger than what was previously measured in hall A ($\phi_A = 16.2_2 \cdot 10^{-6}$ cm$^{-2}$ s$^{-1}$) [7].

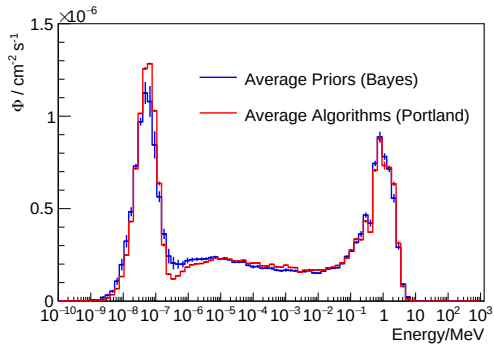

**Figure 5:** Unfolded neutron flux energy spectrum obtained from the average rates in phase 3 (HENSA-v2022, 168 life-days). Red: average from different priors (Bayesian algorithm). Blue: average from different methods (Portland concrete). The error-bars are the standard deviations.

## 6. Summary and future work

The neutron flux has been being measured in hall B of the LSC for more than two years. Preliminary results from the analysis of the neutron rates in the firsts two years of measurement have been reported. The neutron rates appeared to be almost stable within the statistical uncertainties. Semi-independent methodologies providing consistent results have been used. Future plans include the assessment of potential correlations with the environmental variables.

Preliminary unfolding results up to 20 MeV have been reported. The effect of using different priors have been evaluated. Different unfolding algorithms provided consistent results. The total magnitude of the neutron flux has found to be larger than what was previously measured in hall A. This highlights the dependence of the neutron flux on the local environment.

## Acknowledgments

This work has been supported by the Spanish Grants PID2019-104714GB-C21 & C2, FPA2017-83946-C2-1-P & C2-2-P and RTI2018-098868-B-I00 (MCIU/AEI/FEDER), the Generalitat Valenciana Grant PROMETEO/2019/007, the European Union (EU) ChETEC-INFRA 101008324 and the MRR co-financed by the Generalitat de Catalunya and the EU. We are grateful to the LSC for hosting HENSA and for the support received during the measurement campaign.